\documentstyle[12pt]{article}

\begin{document}
\begin{titlepage}
\begin{flushright}\vbox{\begin{tabular}{c}
           TIFR/TH/96-48\\
           August, 1996\\
           hep-ph/yymmdd\\
\end{tabular}}\end{flushright}
\begin{center}
   {\large \bf
      Colour-octet Contributions to $x_{\scriptscriptstyle F}$ Distributions\\
      in $J/\psi$ Hadro-production at Fixed Target Energies.}
\end{center}
\bigskip
\begin{center}
   {Sourendu Gupta and K.\ Sridhar.\\
    Theory Group, Tata Institute of Fundamental Research,\\
    Homi Bhabha Road, Bombay 400005, India.}
\end{center}
\bigskip
\begin{abstract}
We study $J/\psi$ production at fixed-target energies, including the
colour-octet production mechanisms predicted by NRQCD. In an earlier
paper, we found that
the octet components were crucial in understanding $p_{\scriptscriptstyle
T}$-integrated forward ($x_{\scriptscriptstyle F} > 0$) cross-sections.
Here we make a detailed comparison of the theoretical predictions with
measured $x_{\scriptscriptstyle F}$ distributions from fixed-target
experiments. We find that the model predictions agree well with data.
Taking into account higher orders in NRQCD, we argue that $\sigma(\chi_1)/
\sigma(\chi_2)=0.6$, in excellent agreement with data. Similiar arguments
also predict that $\sigma(\chi)/\sigma(J/\psi)$ is roughly equal in
photo- and hadro-production.
\end{abstract}
\end{titlepage}

\section{Introduction}

A large volume of data now exists on the hadro-production of $J/\psi$ at
fixed target energies--- both for the total cross section and the
$x_{\scriptscriptstyle F}$ distribution. In particular,
several new experiments have taken high statistics data with pion or
proton beams. Since a promising new model for heavy-quarkonium production
is now available, it is useful to perform an analysis in the context of
this model. This is the purpose of this paper.

The physics of heavy quarks can be expected to be dominated by its (large)
mass, $m$, and hence computable in perturbative QCD. However, for quarkonium
systems the scales $mv$ (relative momentum) and $mv^2$ (energy) are also
important, since the quarks are expected to be non-relativistic
in the rest frame of the quarkonium. An organised way to include the effects
of the dimensionless parameter $v$ (relative velocity) is provided by an
effective low-energy field theory called non-relativistic QCD (NRQCD)
\cite{caswell}. 

A cutoff of the order of $m$ is introduced \cite{bbl} into the QCD action
to remove all states of momenta larger than $m$. Their effects are taken
into account through new couplings, which are local since the excluded states
are relativistic. Beyond the leading order in $1/m$ the effective theory is
non-renormalisable. This cut-off theory is block-diagonalised to decouple
the heavy-quark degrees of freedom, leaving a non-relativistic Schr\" odinger
field theory. Consequently, states of a quark anti-quark pair can be
specified by the spin, orbital and total angular momentum (${}^{2S+1}L_J$ in
spectroscopic notation), a radial quantum number and the colour representation.
Gluons emitted by quarks can be specified in a multipole expansion. The
formalism of NRQCD has been successfully applied to first principles lattice
computations of quarkonium spectra \cite{lattice}.

In this framework, a $c\bar c$ pair is created by a short-distance process,
and it binds into a quarkonium state over scales that are longer by powers
of $1/v$. Assuming factorisation, the cross-section for the production of a
meson $H$ can be written as
\begin{equation}
   \sigma(H)\;=\;{\rm Im}\,\sum_{n=\{\alpha,S,L,J\}} {F_n\over m^{d_n-4}}
       \langle{\cal O}^H_\alpha({}^{2S+1}L_J)\rangle
\label{e1}\end{equation}
where $F_n$'s are short distance matrix elements and ${\cal O}_n$ are local
4-fermion operators, of naive dimension $d_n$, describing the long-distance
physics (the angular brackets denote a vacuum expectation value).
The cutoff-dependence of $F_n$ is compensated by that of the long-distance
matrix elements. The colour index $\alpha$ can run over singlet as well as
octet representations.

The short-distance coefficients $F_n$ have the usual perturbation expansion
in powers of $\alpha_{\scriptscriptstyle S}$. The scaling of the matrix
element $\langle{\cal O}^H_\alpha({}^{2S+1}L_J)\rangle$ with $v$ is determined
by the (rotational) tensor representation of the gluonic transition connecting
the state of the produced pair to the target hadron. Consequently, the sum
in eq.\ (\ref{e1}) is a double power series expansion in $v$ and
$\alpha_{\scriptscriptstyle S}$. For bottomonium states, $v^2$ is small
compared to $\alpha_{\scriptscriptstyle S}(m^2)$ and hence colour octet
contributions
are not very significant except when the singlet terms vanish. As a result,
the expansion is close to the normal perturbative expansion. For charmonium
states, a numerical coincidence, $v^2\sim\alpha_{\scriptscriptstyle S}(m^2)$,
makes the double expansion more complicated.

The importance of colour octet pieces were first seen \cite{jpsi} in the
phenomenology of $P$-state charmonium production at large
$p_{\scriptscriptstyle T}$ in Tevatron data \cite{cdf}. These processes
do not have a consistent description in terms of colour singlet operators
only \cite{bbl2}. However, even for the direct production of $S$-states
such as the $J/\psi$ or $\psi'$, where the colour singlet components give
the leading contribution in $v$, the inclusion of sub-leading octet states
was seen to be necessary for phenomenological reasons \cite{brfl}. This
can be understood in terms of the numerical coincidence of $v$ and
$\alpha_{\scriptscriptstyle S}$ mentioned earlier.

Since the long-distance matrix elements are not calculable, and the octet
matrix elements are not easily written in terms of decays, the Tevatron
data must be used to fix these unknown parameters \cite{cho1}.
As a consequence, quantitative test of this formalism can come only from
other experiments. Fixed-target experiments provide such tests, because
colour-singlet contributions are expected to be negligible for
photo-production and hadro-production of $J/\psi$ in appropriate kinematic
regions. A new linear combination of octet matrix elements appears here.
In Ref.~\cite{flem2} this combination has been fixed through an analysis
of elastic photo-production data.

In an earlier paper \cite{ours}, we had considered data on hadro-production
of $J/\psi$ in proton-nucleon and pion-nucleon collisions at centre of mass
energies, $\sqrt S$, upto about 60 GeV. We found that the data on forward
($x_{\scriptscriptstyle F}>0$) integrated cross sections are well described
by the inclusion of octet components. These are parameter-free predictions,
because the values of all the required non-perturbative matrix elements are
derived from other experiments. In this paper, we carry out a more detailed
investigation of the $x_{\scriptscriptstyle F}$ and $y$ distributions of
$J/\psi$ production for the same fixed-target energies. These are given in
Sections 2 and 3.

The octet model predictions for $J/\psi$ production in
fixed-target experiments have also been considered in Ref.~\cite{br,tv}.
In Ref.~\cite{tv} it is claimed that the ratio of directly produced
$J/\psi$ to $\chi$ states is not in agreement with data. This is disputed
in Ref.~\cite{br}. We address this question in Section 4. Our results are
in agreement with those of Ref.~\cite{br}.

\section{Cross Sections}
The $x_{\scriptscriptstyle F}$ distribution for inclusive production of
$J/\psi$ can be written as a sum over all possible channels in the form
\begin{equation}
  {d\sigma_{J/\psi}\over dx_{\scriptscriptstyle F}}\;=\;
     \sum_{n=\{\alpha,S,L,J,H\}}
       {d\sigma^n\over dx_{\scriptscriptstyle F}}
           \langle{\cal O}_\alpha^H({}^{2S+1}L_J)\rangle BR(H\to J/\psi).\\
\label{sigma}\end{equation}
A channel is specified by the rotational and colour quantum
numbers of the $c\bar c$ state, and the colour singlet meson it goes into.
For both colour singlet and octet channels, the strong interaction effects
required to bind a pair into a physical meson are specified by the matrix
element $\langle{\cal O}\rangle$. Subsequently, this meson decays into
the $J/\psi$ with a branching ratio specified by $BR$. The formul\ae{} for
the channel cross sections \cite{flem1}, matrix elements and branching 
ratios are collected in Table \ref{tb.inputs} for all the relevant channels.

\begin{table}\begin{center}
  \begin{tabular}{|c|c|c|c|c|c|}  \hline
  $H$ & $\alpha$ & ${}^{2S+1}L_J$
     & ${\displaystyle d\sigma\over\displaystyle dx_{\scriptscriptstyle F}}$
     & $\langle{\cal O}_\alpha^H({}^{2S+1}L_J)\rangle$
     & $BR$ (\%)\\
  \hline
  $J/\psi$ & 8 & ${}^1S_0$, ${}^3P_0$, ${}^3P_2$
     &${\cal L}_g {\displaystyle5\alpha_{\scriptscriptstyle S}^2\pi^3\over\displaystyle48m^5}$
     & $0.020\pm0.001{\rm\ GeV}^3$
     & 100 \\
  $J/\psi$ & 8 & ${}^3S_1$
     &${\cal L}_q {\displaystyle\alpha_{\scriptscriptstyle S}^2\pi^3\over\displaystyle54m^5}$
     & $0.0066\pm0.0021{\rm\ GeV}^3$
     & 100 \\
  $\chi_0$ & 1 & ${}^3P_0$
     &${\cal L}_g {\displaystyle\alpha_{\scriptscriptstyle S}^2\pi^3\over\displaystyle48m^7}$
     & $0.32\pm0.04{\rm\ GeV}^5$
     & $0.66\pm0.18$ \\
  $\chi_0$ & 8 & ${}^3S_1$
     &${\cal L}_q {\displaystyle\alpha_{\scriptscriptstyle S}^2\pi^3\over\displaystyle54m^5}$
     &
     & $0.66\pm0.18$ \\
  $\chi_1$ & 8 & ${}^3S_1$
     &${\cal L}_q {\displaystyle\alpha_{\scriptscriptstyle S}^2\pi^3\over\displaystyle54m^5}$
     & $0.0098\pm0.0013{\rm\ GeV}^3$
     & $27.3\pm1.6$ \\
  $\chi_2$ & 1 & ${}^3P_2$
     &${\cal L}_g {\displaystyle\alpha_{\scriptscriptstyle S}^2\pi^3\over\displaystyle180m^7}$
     & $0.32\pm0.04{\rm\ GeV}^5$
     & $13.5\pm1.1$ \\
  $\chi_2$ & 8 & ${}^3S_1$
     &${\cal L}_q {\displaystyle\alpha_{\scriptscriptstyle S}^2\pi^3\over\displaystyle54m^5}$
     &
     & $13.5\pm1.1$ \\
  \hline
  \end{tabular}\end{center}
  \caption[dummy]{The cross sections, non-perturbative matrix
     elements and branching ratios into the $J/\psi$ for each channel
     (specified by the colour representation $\alpha$, rotational quantum
     numbers and the primary meson $H$) relevant to the inclusive $J/\psi$
     cross section of eq.\ (\ref{sigma}). The gluon and quark luminosities
     are defined in eq.\ (\ref{lumin}). The value of the matrix element
     quoted in the first line is for the linear combination of
     eq.\ (\ref{psi}).}
\label{tb.inputs}\end{table}

The matrix element appearing in the octet ${}^3S_1$ channel with the gluon
initiated part of the reaction is the combination
\begin{equation}
  \Theta\;=\;
          \left\langle{\cal O}^{J/\psi}_8({}^1S_0)\right\rangle
    +{\displaystyle3\over\displaystyle m^2}
          \left\langle{\cal O}^{J/\psi}_8({}^3P_0)\right\rangle
    +{\displaystyle4\over\displaystyle5m^2}
          \left\langle{\cal O}^{J/\psi}_8({}^3P_2)\right\rangle.
\label{psi}\end{equation}
Precisely this combination appears in the cross section for photo-production
of the $J/\psi$, and has been fitted to the relevant data \cite{flem2}. The
two matrix elements in the singlet ${}^3P_J$ state are related to the
derivative of the wavefunction of $\chi_J$ at the origin by
\begin{equation}
   \left\langle{\cal O}^{\chi_J}_1({}^3P_J)\right\rangle
      \;=\;{9(2J+1)\over2\pi}|R'(0)|^2.
\label{deriv}\end{equation}
It has been fixed by decay rates \cite{mangano}. Each of the three octet
${}^3P_J$ matrix elements have been extracted from hadro-production
rates at the Tevatron. We use the
tabulation of \cite{cho2}, along with the relation
\begin{equation}
   \left\langle{\cal O}^{\chi_J}_8({}^3S_1)\right\rangle\;=\;
       (2J+1)\left\langle{\cal O}^{\chi_0}_8({}^3S_1)\right\rangle.
\label{jscale}\end{equation}
The value of this matrix element for $J=1$ is shown in Table \ref{tb.inputs}.

The gluon and quark luminosities, ${\cal L}_g$ and ${\cal L}_q$ respectively,
are defined in terms of the relevant parton distributions in the projectile
(P) and target (T), through the formul\ae{}---
\begin{equation}\begin{array}{rl}
   {\cal L}_g\;=&\; {\displaystyle1\over\displaystyle\sqrt{x_{\scriptscriptstyle F}^2+4\tau}}
       \,x_+g_{\scriptscriptstyle P}(x_+)
       \,x_-g_{\scriptscriptstyle T}(x_-),\\
   {\cal L}_q\;=&\; {\displaystyle1\over\displaystyle\sqrt{x_{\scriptscriptstyle F}^2+4\tau}}
       \left[\sum_f \,x_+q^f_{\scriptscriptstyle P}(x_+)
       \,x_-\bar q^f_{\scriptscriptstyle T}(x_-) + (P\leftrightarrow T)\right],
\end{array}\label{lumin}\end{equation}
where the sum in the definition of ${\cal L}_q$ is over flavours. We have
used the notation
\begin{equation}
   x_\pm\;=\; {1\over2}(\sqrt{x_{\scriptscriptstyle F}^2+4\tau}\pm x_{\scriptscriptstyle F}).
\label{xdef}\end{equation}
We also discuss briefly existing data on rapidity ($y$) distributions of the
$J/\psi$. These can be easily obtained from the $x_{\scriptscriptstyle F}$
distribution by multiplying the latter with the Jacobian factor
$\sqrt{x_{\scriptscriptstyle F}^2+4\tau}$ for the transformation between
$x_{\scriptscriptstyle F}$ and $y$. It is clear that the dependence of
the cross sections on $x_{\scriptscriptstyle F}$ or $y$ comes entirely from
these luminosities.

Before discussing our computations and the comparison with data, we would
like to note that the colour octet model contains many parameters in the
form of non-perturbative matrix elements. Our approach here is to fix these
values from different experiments and use the fixed target hadro-production
data as a test of the phenomenological viability of the model.

\section{Results}

\begin{figure}
\vskip14truecm
\includegraphics{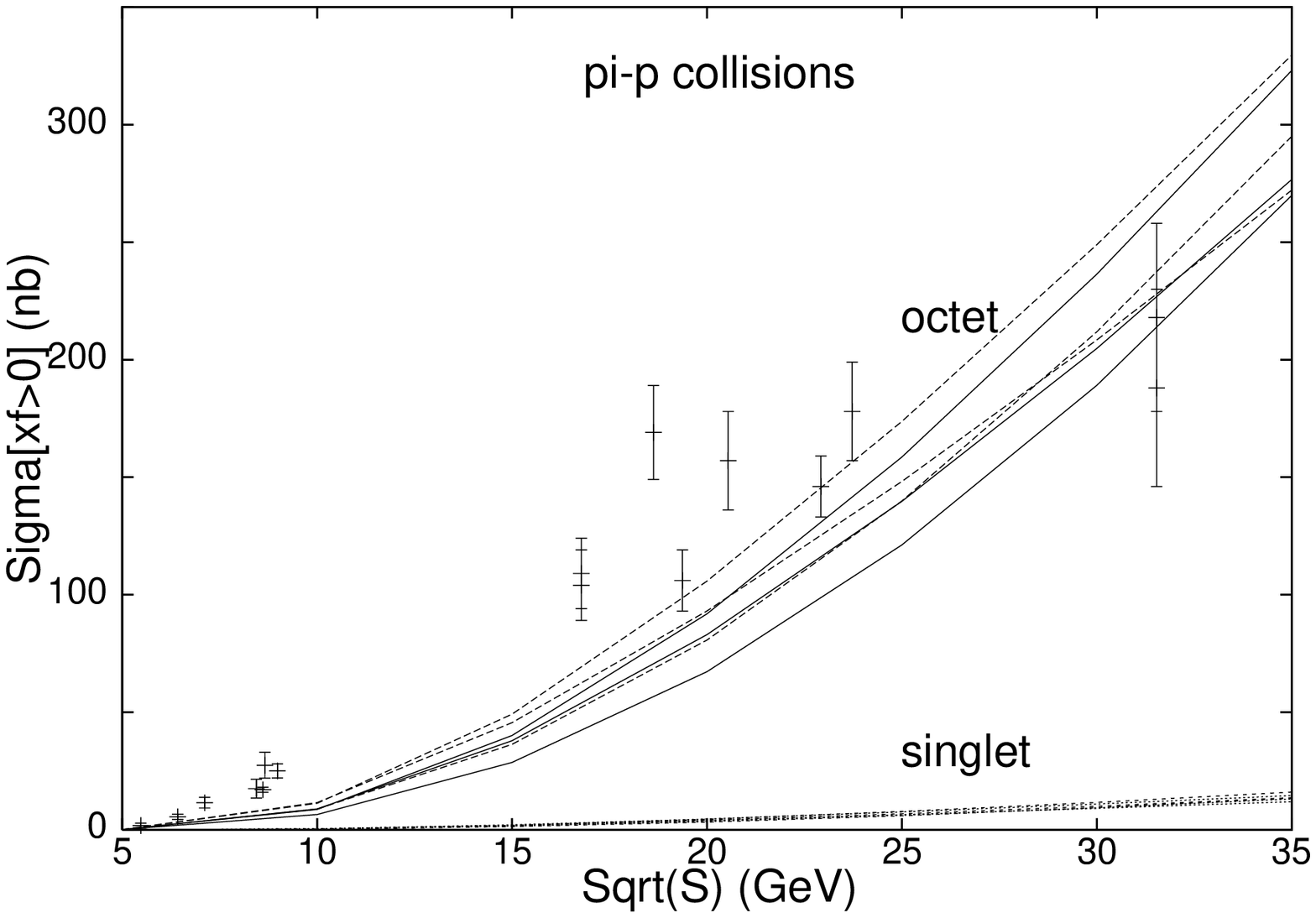}
\includegraphics{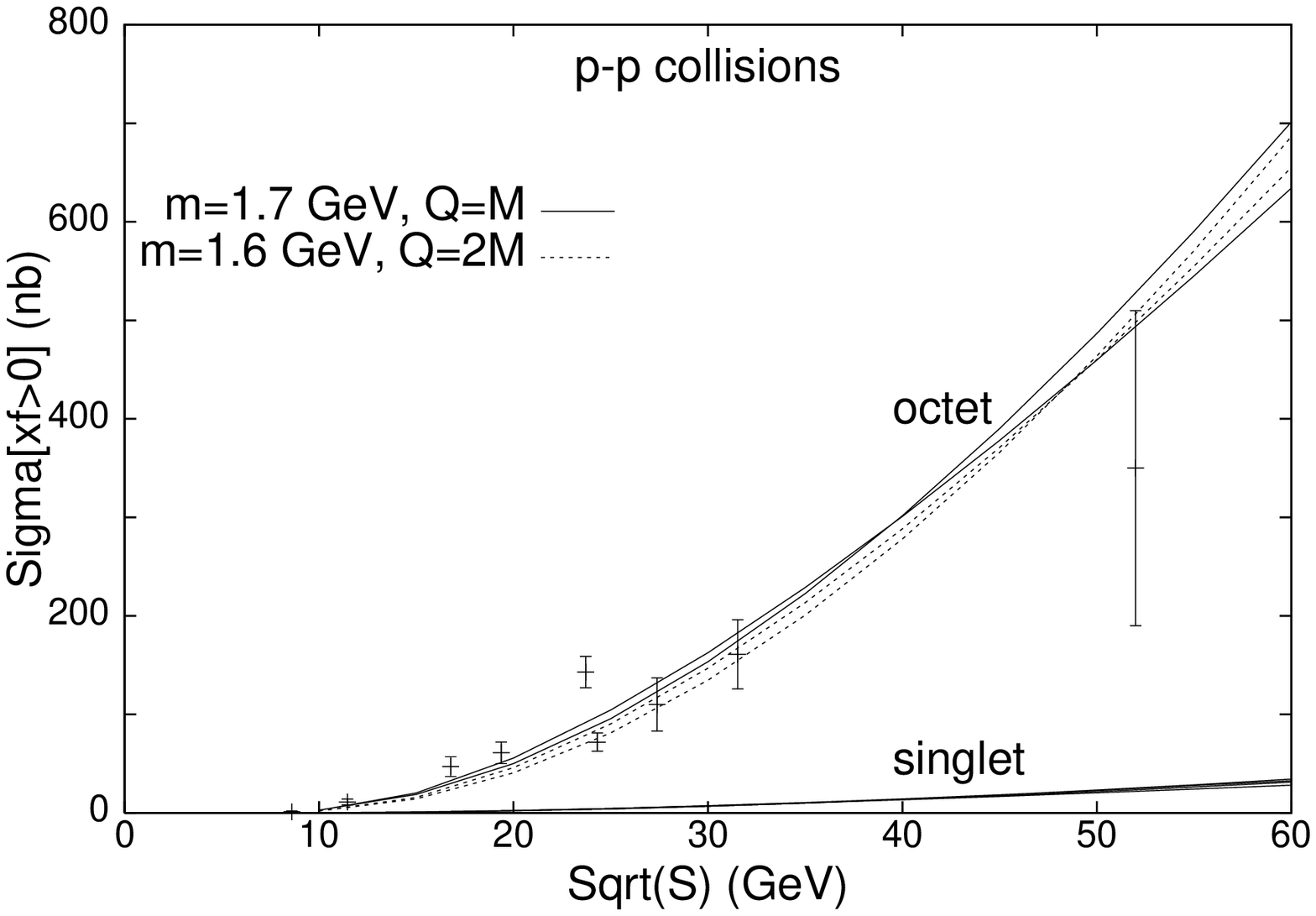}
\caption[dummy]{The colour-octet model predictions for integrated forward
  $J/\psi$ hadroproduction cross sections as a function of the CM energy.
  For $pp$ collisions, the two curves for each choice of $m$ and scale $Q$
  are for the structure functions MRS ${\rm D}-'$ and GRV LO. For $\pi p$
  collisions the three sets of structure functions are
  MRS ${\rm D}-'$ for proton and SMRS 1 for $\pi$, MRS ${\rm D}-'$
  for proton and SMRS 3 for $\pi$, GRV LO for both. Note that the
  colour-singlet model predictions lie far below the data.}
\label{fg.tot}\end{figure}

Before embarking on the computation of cross sections there are the usual
choices for QCD inputs to be fixed up. From open charm production, the
limits on the charm quark mass are $1.2 {\rm\ GeV}\le m \le 1.8$ GeV
\cite{nason}. In an earlier calculation of the total forward cross section,
$\sigma(x_{\scriptscriptstyle F}>0)$ \cite{ours}, we had varied $m$ between
$1.6$ GeV and $1.7$ GeV to obtain an agreement between the data and model
computations. The QCD scale, $Q$, was varied between $M$ and $2M$ (where $M$
is the mass of the $J/\psi$). For these calculations we had used the
${\rm MRS\ D}-'$ and the GRV LO sets of parton densities for the proton,
and the SMRS 1, SMRS 3 and GRV densities for the pion. All the sets were
taken from the PDFLIB package \cite{pdflib}. We continue with these choices
here.

The total forward cross sections are shown in Figure \ref{fg.tot}. We would
like to point out the following features---
\begin{itemize}
\item
   The colour singlet model predictions lie far below the data. The
   addition of the octet channels is necessary for the phenomenology.
\item
   For $pp$ collisions, all the data lie on the curves shown,
   with the exception of the data from \cite{pp300}. The total forward cross
   section with a 300 GeV proton beam, reported in \cite{pp300}, is $143\pm17$
   nb. This is much higher than the corresponding number, $71.8\pm9.3$,
   reported by \cite{ua6} from a 315 GeV proton beam. It would be useful to
   have a third experiment to sort out this mismatch.
\item
   For $\pi p$ collisions the data are very scattered. More data would
   be very welcome. The structure functions for $\pi$ are also rather badly
   known, and this shows up as a bigger spread in the model prediction.
   Further experiments on high-mass lepton pair production with pion beams
   could reduce this uncertainty.
\item
   The data at the highest energies (ISR) seem to prefer a smaller value of
   $m$. We discuss this point later.
\end{itemize}

\begin{figure}
\vskip14truecm
\includegraphics{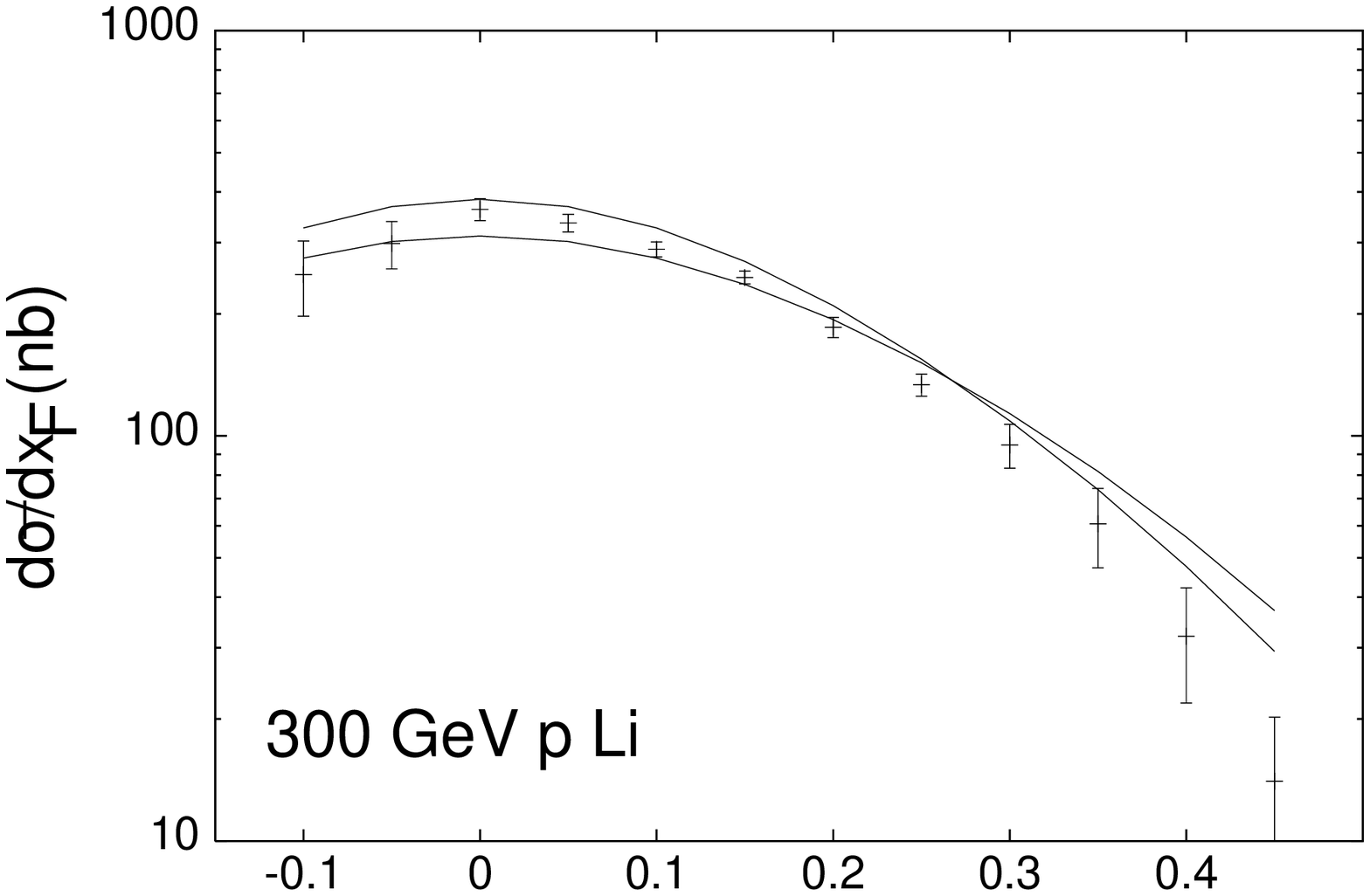}
\includegraphics{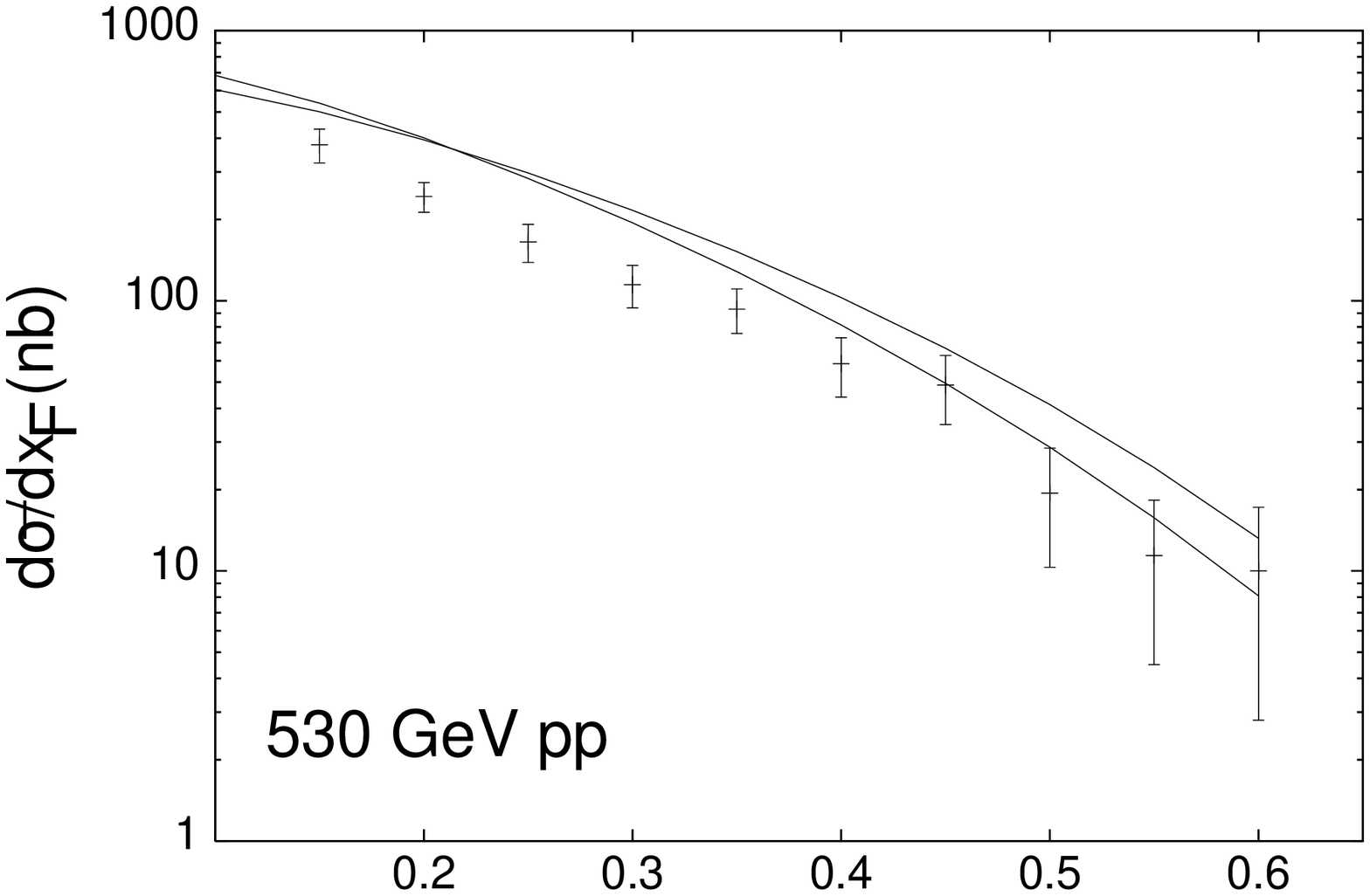}
\includegraphics{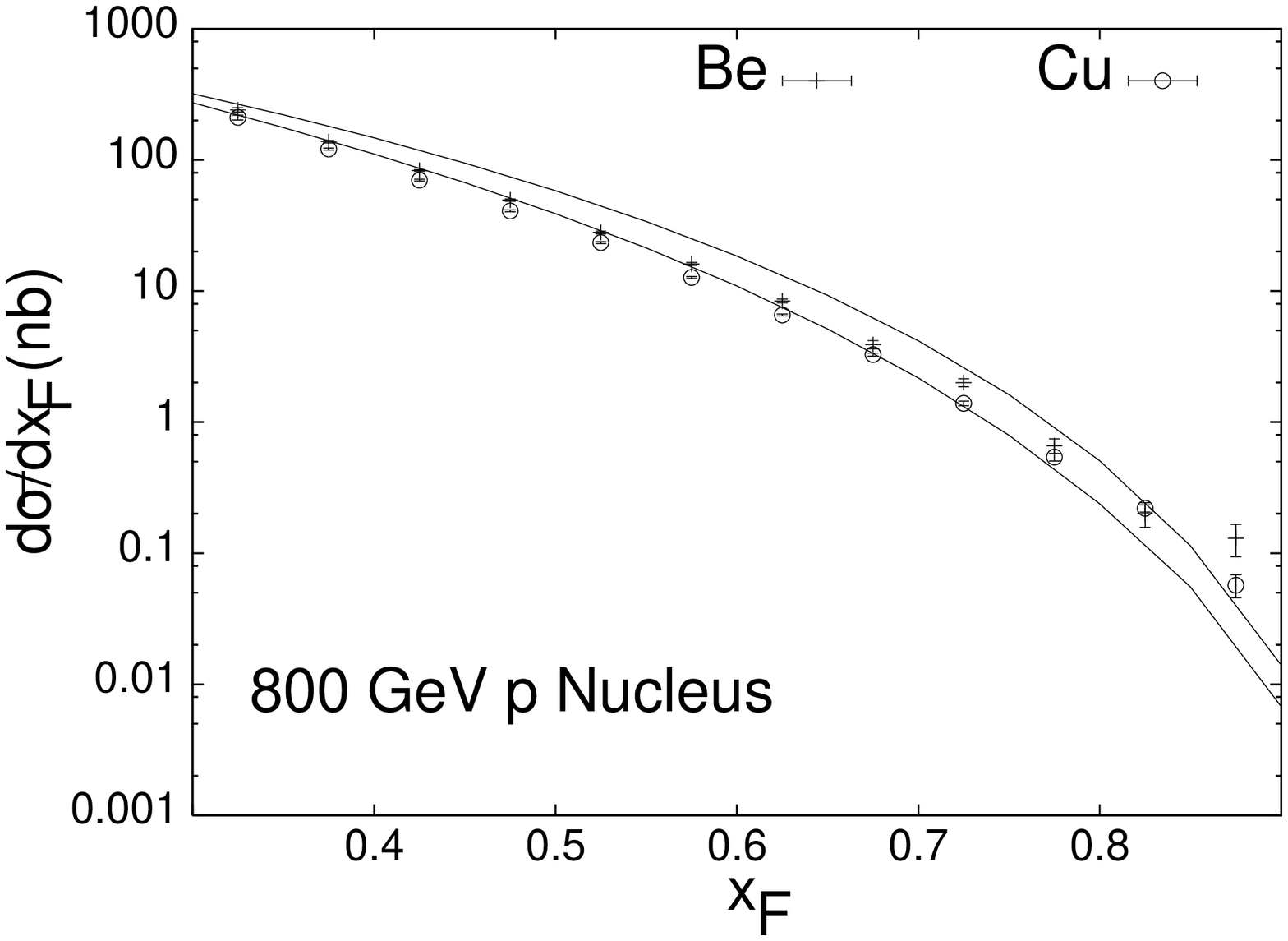}
\caption[dummy]{The colour-octet model predictions for the cross section
  differential in $x_{\scriptscriptstyle F}$ for $J/\psi$ production in $pp$
  collisions for different beam energies. The two curves are for the parton
  density sets ${\rm MRS\ D}-'$ and GRV LO and $m=1.7$ GeV and $Q=M$.}
\label{fg.pp}\end{figure}

In Figure \ref{fg.pp}, we compare the model with data on $x_{\scriptscriptstyle
F}$ distributions from three different
$pp$ experiments \cite{pp300,pp550,pp800}. There is extremely good agreement
for proton beam energy of 800 GeV over the 4 orders of magnitude spanned by
the data \cite{pp800}. Since the data sample contains $2.45\times10^5$ events,
the quoted errors are very small and the agreement is very significant. Data
at other beam energies are also well described. The $x_{\scriptscriptstyle F}$
distribution at 300 GeV energy \cite{pp300} has been normalised by the ratio
of their measured value of $\sigma(x_{\scriptscriptstyle F}>0)$ and the
model prediction. There is no statistically significant evidence of a
systematic deviation from the model predictions at large
$x_{\scriptscriptstyle F}$.
At large values of $x_{\scriptscriptstyle F}$ the 300 GeV data may lie slightly
below the model (but with statistically insignificant deviation). On the other
hand, data in the last $x_{\scriptscriptstyle F}$ bin for the
800 GeV set lies a little above the prediction. Note also that only the
statistical errors on the data are shown in the figures.

\begin{figure}
\vskip14truecm
\includegraphics{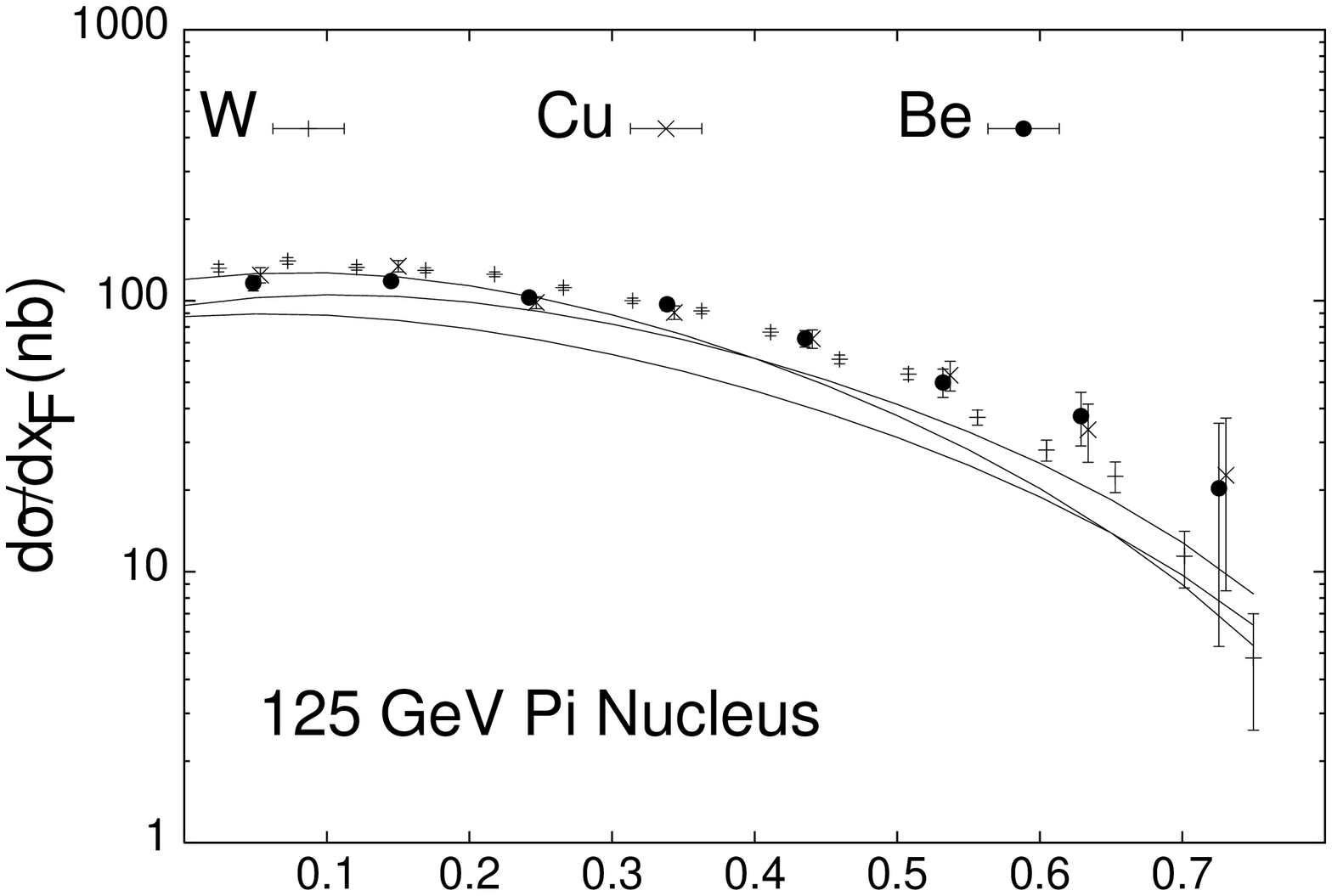}
\includegraphics{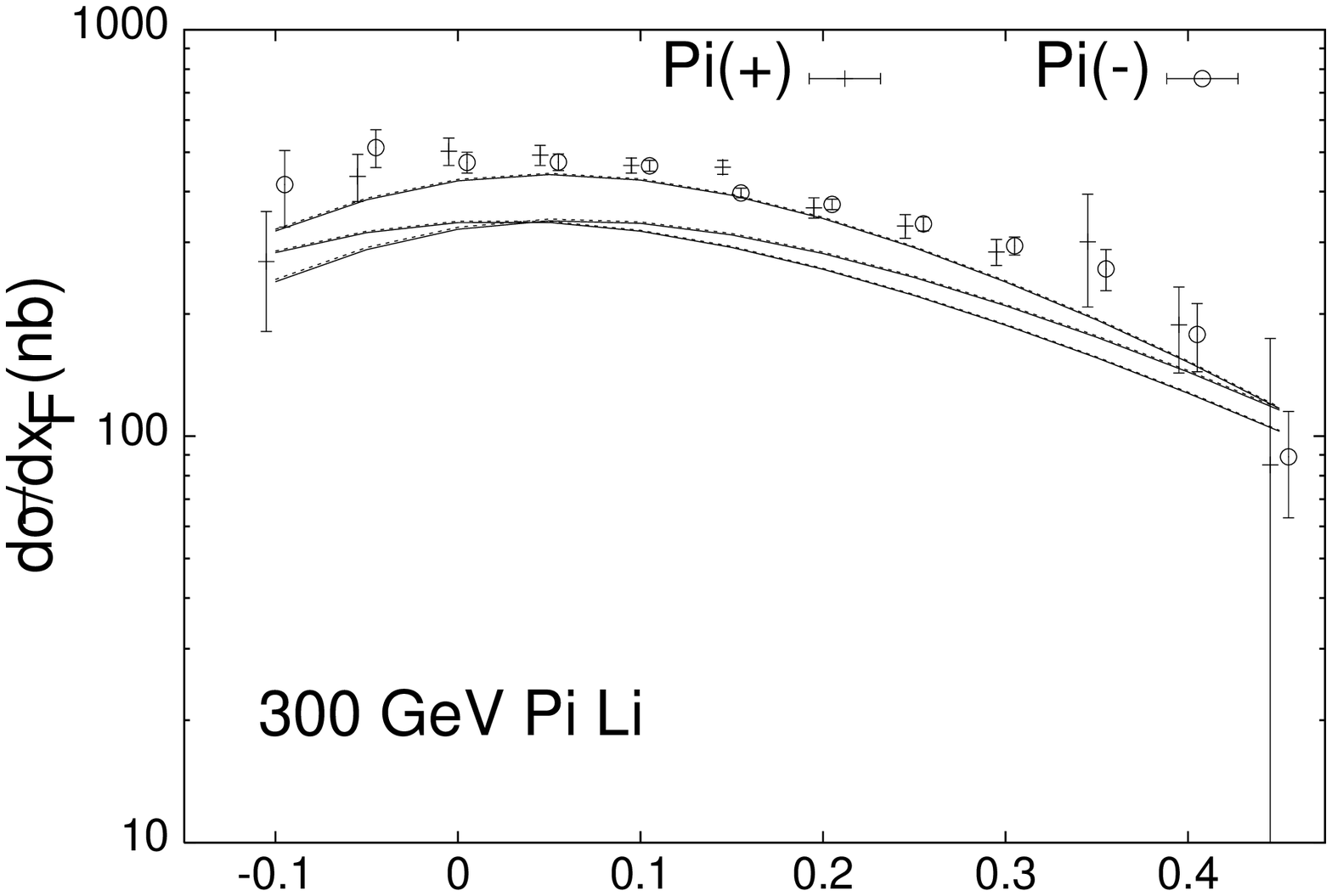}
\includegraphics{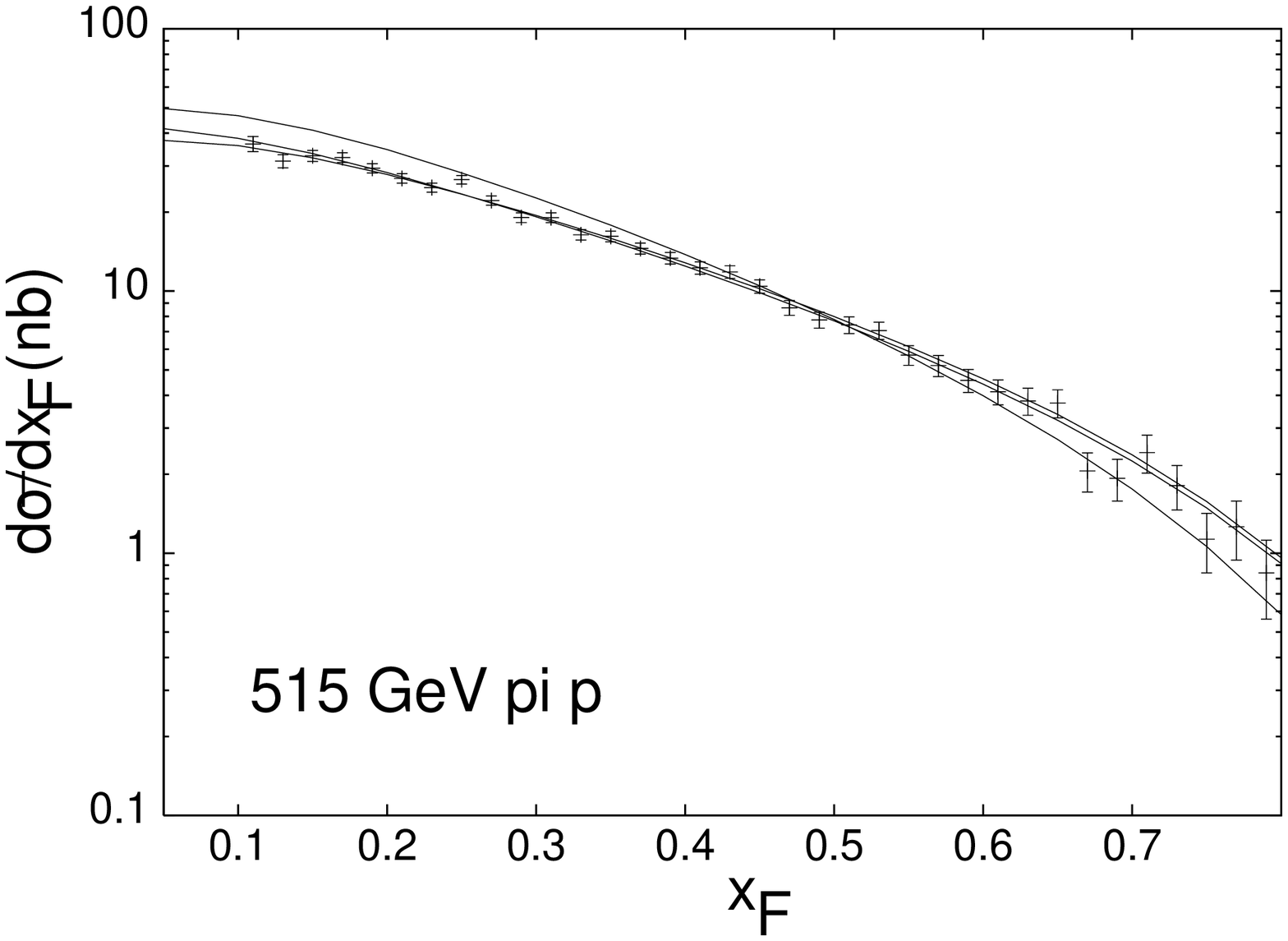}
\caption[dummy]{The colour-octet model predictions for the cross section
  differential in $x_{\scriptscriptstyle F}$ for $J/\psi$ production in $\pi p$
  collisions for different beam energies. The two curves are for the parton
  density sets ${\rm MRS\ D}-'$ and GRV LO and $m=1.7$ GeV and $Q=M$. For
  visibility, the $\pi^+$ and $\pi^-$ data for beam energy 300 GeV have been
  offset equally to either side of the true $x_{\scriptscriptstyle F}$ value.
  This is also done for the Cu and Be data at 125 GeV.}
\label{fg.pip}\end{figure}

Figure \ref{fg.pip}, shows the $x_{\scriptscriptstyle F}$ distributions
in $\pi-p$ collisions. The data are taken from three different experiments
\cite{pp300,pip125,pip515}. At the highest beam energy of 515 GeV, the
data agrees extremely well with the model predictions both in magnitude
and shape. At lower beam energies the shape is reproduced extremely well,
although the agreement in magnitude is less perfect. It is not
possible to decide whether this is due to nuclear effects or the pion
structure functions, or even experimental uncertainties in the normalisation
(see Figure \ref{fg.tot}). More data would help in deciding between these
possibilities.

There is a small body of literature on cross sections and distributions
in $\bar pp$ collisions. The data on total forward cross sections
is even more scattered than the pion data. Also, the number of $J/\psi$
events are rather low, and hence lead to larger errors. For these reasons
we have not attempted a quantitative study of this data.

\begin{figure}
\vskip14truecm
\includegraphics{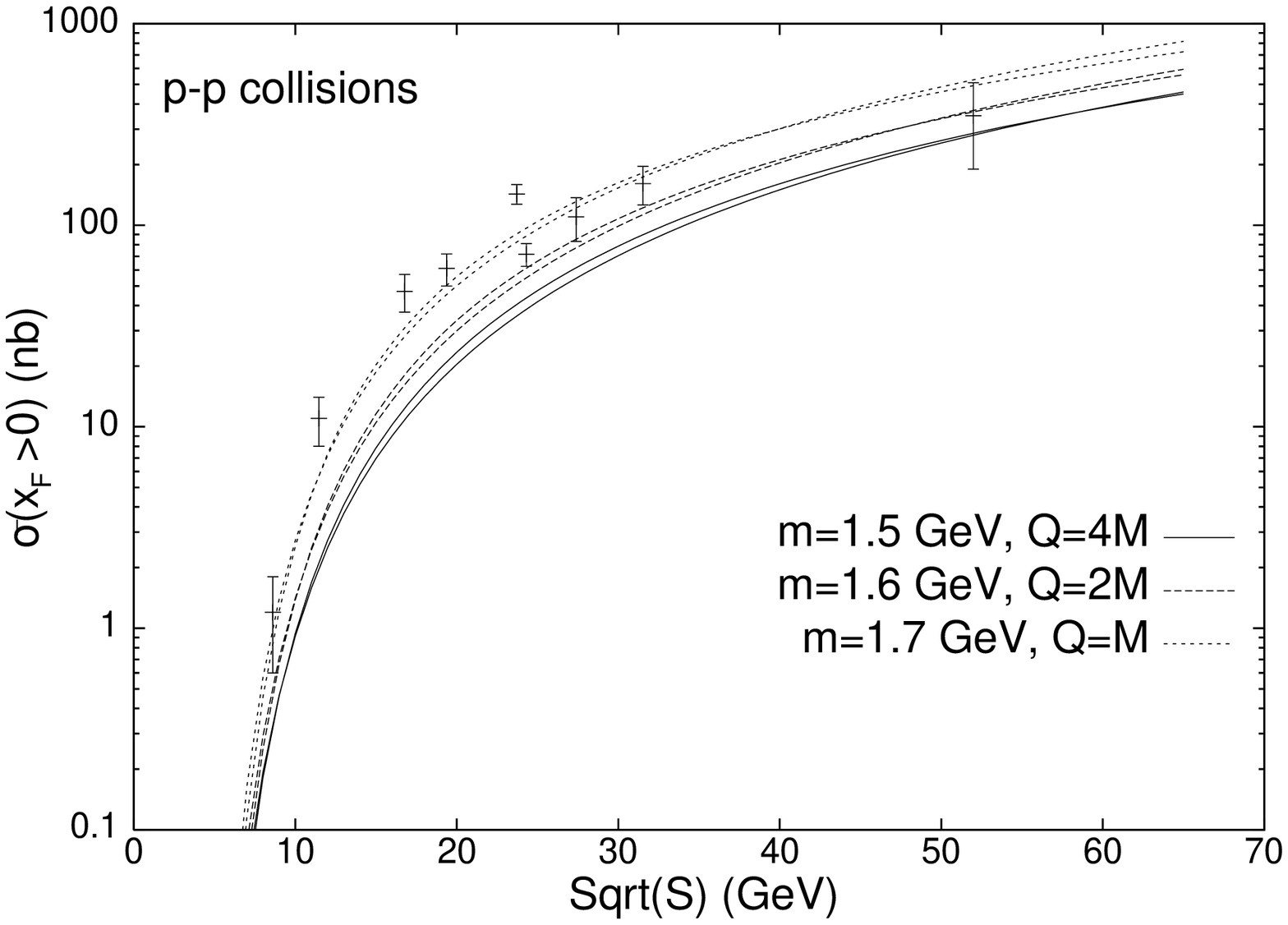}
\includegraphics{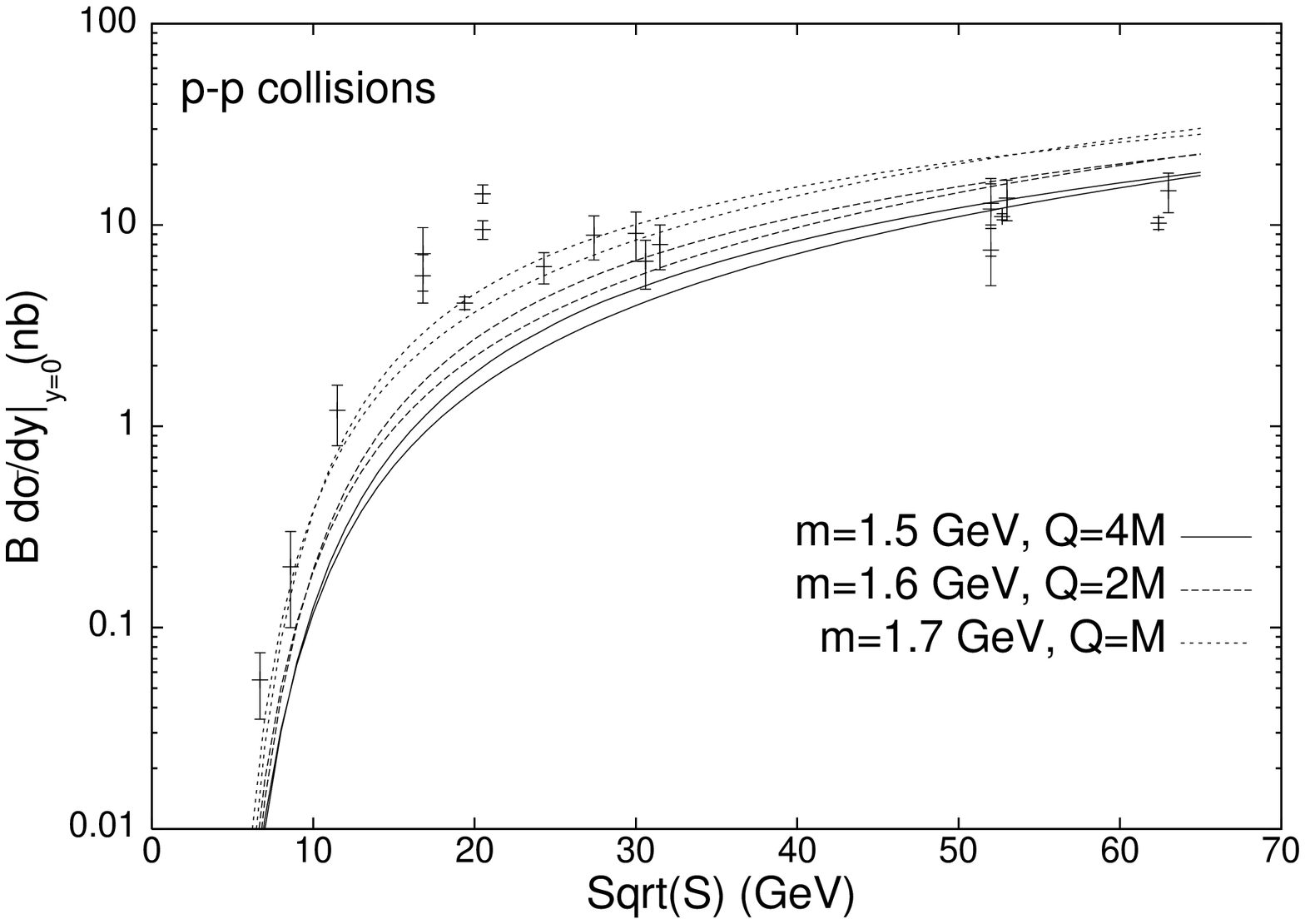}
\caption[dummy]{$\sigma(x_{\scriptscriptstyle F}>0)$ for $J/\psi$ production
   in $pp$ collisions (upper panel) and the differential cross section
   $Bd\sigma/dy$ at y=0 (lower panel), shown as a function of $\sqrt S$.
   For the model calculation we have assumed that the branching ratio for
   the decay of the $J/\psi$ into leptons ($B$) is 5.97\%.}
\label{fg.ydep}\end{figure}

Since a large amount of data exists for the differential cross section
$d\sigma/dy$ at $y=0$ in $pp$ collisions, we have also tested the model
against this. The data goes from values of $\sqrt S$ near threshold upto
ISR energies. The comparison of model and data are shown in Figure
\ref{fg.ydep}. It is clear that the data on the rapidity distribution is
more scattered than that on total forward cross sections. In several
cases experiments at the same value of $\sqrt S$ yield values for the
distribution which are not in agreement even when errors are taken into
account. Nevertheless, a reasonable overall agreement of the model with
the data can be seen.

It is interesting to note, however, that the low $\sqrt S$ data prefer a
larger value of the charm quark mass, $m$, and a smaller scale $Q$ than the
ISR data. As shown in the figure, this trend holds for both the total
forward cross section and the rapidity distribution. A similiar observation
used to be made for single inclusive charm production, before higher order
perturbative calculations resolved this problem \cite{nde}. The essential
physics is that higher orders are important near threshold because
an off-shell heavy quark can be thrown on-shell by soft radiation. It seems
reasonable to expect that similiar effects are at work here. Further study
would call for inclusion of higher order perturbative effects.

Very naively one may have thought of tuning $m$ and $Q$ in order to produce
the best results. However, since such a tuning would depend on $\sqrt S$,
it is clearly not the correct way to proceed. In fact, as is now well-known,
such sensitive dependence on the scale is an artifact of leading order
perturbative QCD calculations. Once higher orders are computed, it is
possible to remove this strong dependence on $Q$ and then fix the best
values for $m$ \cite{nason}. Hence, we may take the joint dependence of
the cross sections on $m$ and $Q$ as an estimate of the importance of higher
order perturbative QCD effects.

\begin{table}\begin{center}
  \begin{tabular}{|c|c|c|}  \hline
     Source of uncertainty & $pp$ & $\pi p$\\ \hline
  
     Parton density sets &
        20\% & 29\% \\

     Scale uncertainties &
        91\% & 96\% \\

     Non-perturbative matrix elements &
        5\% & 5\% \\
  \hline
   \end{tabular}\end{center}
   \caption[dummy]{Estimates of the magnitude of various theoretical
      uncertainties in the colour octet model for $J/\psi$ production
      at $\sqrt S=19.4$ GeV. These are obtained from a calculation of
      $d\sigma/dx_{\scriptscriptstyle F}$ at $x_{\scriptscriptstyle F}
      =0$.}
\label{tb.err}\end{table}

The magnitudes of the various theoretical uncertainties are summarised
in Table \ref{tb.err}. The errors on the individual non-perturbative
matrix elements are quoted in Table \ref{tb.inputs}. The dominant channel
in eq.\ (\ref{sigma}) is the gluon fusion part of the octet $J/\psi$
(the other channels are suppressed either by a branching ratio or by
${\cal L}_q/{\cal L}_q$). As a result, the corresponding matrix element
dominates this source of uncertainty. The error quoted in the table
above is a 1-$\sigma$ limit. Even if it is replaced by a 3-$\sigma$
limit, the uncertainty due to the matrix elements remains the smallest
source of error. Scale uncertainties
in the cross section are a factor of two. As discussed in the
previous paragraph, these may be taken as a rough estimate of the
importance of higher order perturbative QCD effects.

Apart from the model discussed in this paper, another, called the
semi-local duality model (SLD) \cite{sld}, is also used to describe
charmonium cross sections. This model fixes cross sections for
individual charmonium states by comparison with data on the total
cross section at some value of $\sqrt S$. Once this is done, the
results \cite{gavai} for $\sqrt S$ dependence and
$x_{\scriptscriptstyle F}$ are similiar to those reported here.
A distinction can perhaps be made with accurate data at large
$x_{\scriptscriptstyle F}$, since SLD predicts a somewhat harder
$x_{\scriptscriptstyle F}$ distribution. Other distinctions are
possible \cite{br}, and will be discussed elsewhere.

\section{Discussion}

\begin{figure}
\vskip8truecm
\includegraphics{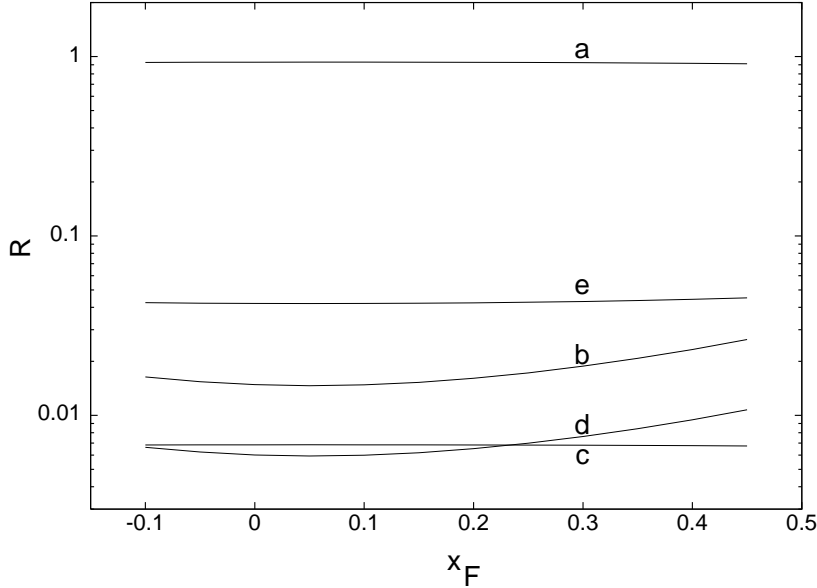}
\caption[dummy]{$R$, the fraction of $d\sigma/dx_{\scriptscriptstyle F}$
   as a function of $x_{\scriptscriptstyle F}$ in to the channels (a)
   $gg\to J/\psi$, (b) $qq\to J/\psi$, (c) $\chi_0$, (d) $\chi_1$ and (e)
   $\chi_2$ production.}
\label{fg.channel}\end{figure}

Next we turn to the ratio $\sigma(\chi)/\sigma(J/\psi)$.
All the hard cross sections in Table \ref{tb.inputs}
are taken to order $\alpha_{\scriptscriptstyle S}$. The non-perturbative
matrix elements in the $\chi_{\scriptscriptstyle J}$ channels are all of
leading order in the velocity--- $v^2$, and there are no contributions
at order $v^4$. The octet terms for $\chi_{\scriptscriptstyle J}$ are
suppressed relative to the singlet terms by ${\cal L}_q/{\cal L}_g$.
The contribution to direct $J/\psi$ production starts at order $v^4$.
Although sub-leading in $v$, it happens to be the channel with the largest
contribution (see Figure \ref{fg.channel}) since it is gluon initiated
and hence comes with a factor of ${\cal L}_g$.
The $\chi_1$ state, which has the largest branching fraction into $J/\psi$
has no singlet contribution and hence is also by ${\cal L}_q/{\cal L}_g$.

From this discussion it is clear that the relative magnitudes of the
parton luminosities,
${\cal L}_g$ and ${\cal L}_q$, are important in determining the dominant
production channels, and may offset the power counting in $v$ to some
extent. At order $v^6$ there are no new contributions to direct $J/\psi$
production, but $\chi_1$ is produced through gluon initial states. Therefore,
we have to consider all production channels upto order $v^6$ to get the
$\chi$ to direct $J/\psi$ ratio. Beyond the order $v^6$ terms, power
counting in $v$ can be used to truncate the NRQCD series at any desired
level of accuracy, since all the dominant ${\cal L}_g$ type terms are
accounted for at order $v^6$.

Enumerating all possible processes at this order, we find that a certain
linear combination, $\Phi$, of $\langle{\cal O}_8^{\chi_J}({}^1S_0)\rangle$,
$\langle{\cal O}_8^{\chi_J}({}^3P_{0,2})\rangle$ and
$\langle{\cal O}_8^{\chi_J}({}^3D_{J'})\rangle$ is required for
the hadro-production cross section \cite{ours2}. These arise equally from
quark and gluon initiated processes and hence the same linear combination
$\Phi$ is associated with both ${\cal L}_g$ and ${\cal L}_q$. Analysis of
the diagrams shows that precisely the same linear combination appears also
in the photo-production cross section.

Two interesting consequences now emerge as predictions of the
NRQCD approach to quarkonium production---
\begin{enumerate}
\item
   There are three families of matrix elements involved in $\Phi$. Each
   scales as $(2J+1)$ (with order $v^2$ corrections) \cite{bbl} where $J$
   is the spin of the $\chi$ state. This gives---
   \begin{equation}
      {\sigma(\chi_1)\over\sigma(\chi_2)}\;=\;\left\{
         \begin{array}{ll}
            {3\over5}&\qquad({\rm NRQCD}),\\
            {0.62\pm0.18\pm0.09}&\qquad({\rm experiments}).\\
         \end{array}\right.
   \label{pred1}\end{equation}
   Within NRQCD, the predicted ratio does not depend on whether the mesons
   are photo- or hadro-produced. Nor does it depend on $\sqrt S$. The
   measurement quoted above comes from a global analysis of hadro-production
   experiments \cite{chiexp}. There are no studies of this ratio in
   photo-production. We would like to urge experimentalists to perform such
   a measurement.
\item
   In photo-production the ratio of $\chi$ and direct $J/\psi$ production
   is determined by the the ratio of the terms in $\Phi$ and $\Theta$ (eq.\ 
   \ref{psi}). As a result this ratio is independent of $\sqrt S$. In
   hadro-production this ratio is altered a little by the contribution of
   the colour-singlet production of $\chi_0$ and $\chi_2$. The ratio increases
   by about $0.05$ (see Figure \ref{fg.channel}). A small $\sqrt S$
   dependence is also to be expected in hadro-production due to the
   $\sqrt S$ dependence of ${\cal L}_q/{\cal L}_g$. This may cause the
   ratio to be slightly larger at small values of $\sqrt S$. A detailed
   numerical treatement will be presented elsewhere \cite{ours2}. Data
   on this ratio from hadro-production yields the value $0.43\pm0.04\pm0.04$
   \cite{chiexp}, almost independent of $\sqrt S$. In photo-production there
   is only the limit $(\chi_1+\chi_2)/(J/\psi) <34\%$ (90\% CL) \cite{na14}.
   Data on this ratio from photo-production experiments would constitute
   an important test of the NRQCD based colour-octet model.
\end{enumerate}

The introduction of these order $v^6$ terms does not change the inclusive
$J/\psi$ cross section because the value of $\Theta$ extracted from
photo-production has to decrease due to the inclusion of $\Phi$. The shape
of the $x_{\scriptscriptstyle F}$ distribution comes from the gluon luminosity.
This is clear from Figure
\ref{fg.channel}, since the ratio of the direct to total $J/\psi$ differential
cross section is constant. Since the order $v^6$ contribution also comes from
a gluon initiated process, and carries a factor ${\cal L}_g$, the shape of the
$x_{\scriptscriptstyle F}$
distribution remains unchanged. The net result of adding all contributions
to the cross section upto order $v^6$ is to change only the $\chi$ to direct
$J/\psi$ ratio. A detailed study will be published later \cite{ours2}.

In summary, the colour-octet model based on NRQCD and an assumption about
factorisation, written to order $\alpha_{\scriptscriptstyle S} v^4$
gives a good phenomenological
description of the total forward $J/\psi$ cross section and the
$x_{\scriptscriptstyle F}$ and $y$ distributions in hadroproduction at fixed
target energies. A small
discrepancy is found in extrapolating the fixed target results to ISR
energies, but these are of a kind that have been earlier seen in
single inclusive charm production and removed there by going to higher
orders in $\alpha_{\scriptscriptstyle S}$.
There is a discrepancy in the ratio of $\chi$ to
direct $J/\psi$ production at order $v^4$. We have argued that this problem can
be solved without changing the results on the cross section by extending
the calculations to order $\alpha_{\scriptscriptstyle S} v^6$. Inclusion
of these order $v^6$ terms leads to two predictions--- that for the
$\chi_1/\chi_2$ ratio is borne out by data, and the other, the rough
equality of the $\chi/J/\psi$ ratio in photo- and hadro-production can
easily be tested in experiments.
We believe that NRQCD provides
a phenomenologically viable model for quarkonium production. Control of the
next orders in the
expansions in $v$ and $\alpha_{\scriptscriptstyle S}$ are, however, necessary.
This work is now in progress \cite{ours2}.

%


\begin{thebibliography}{99}
\bibitem{caswell} 
   W.E.\ Caswell and G.P.\ Lepage, 
  {\sl Phys.\ Lett.\/}, B 167 (1986) 437.
\bibitem{bbl}
   G.\ T.\ Bodwin, E.\ Braaten and G.\ P.\ Lepage,
   {\sl Phys.\ Rev.\/}, D 51 (1995) 1125.
\bibitem{lattice}
   P.\ Lepage {\sl et al.\/}, {\sl Phys.\ Rev.\/}, D 46 (1992) 4052;\\
   C.\ T.\ H.\ Davies {\sl et al.\/}, {\sl Phys.\ Rev.\/}, D 52 (1995) 6519;\\
   C.\ T.\ H.\ Davies {\sl et al.\/}, {\sl Phys.\ Rev.\/}, D 50 (1994) 6963;\\
   C.\ T.\ H.\ Davies {\sl et al.\/}, {\sl Phys.\ Lett.\/}, B 345 (1995) 42;\\
   C.\ T.\ H.\ Davies {\sl et al.\/},
       {\sl Phys.\ Rev.\ Lett.\/}, 73 (1994) 2654.
\bibitem{jpsi}
   E.\ Braaten, M.\ A.\ Doncheski, S.\ Fleming and M.\ Mangano,
     {\sl Phys.\ Lett.\/}, B 333 (1994) 548;\\
   D.\ P.\ Roy and K.\ Sridhar, {\sl Phys.\ Lett.\/}, B 339 (1994) 141;\\
   M.~Cacciari and M.~Greco, {\sl Phys.\ Rev.\ Lett.\/}, 73 (1994) 1586. 
\bibitem{cdf}
   F.\ Abe {\sl et al.\/}, {\sl Phys.\ Rev.\ Lett.\/}, 69 (1992) 3704 and
     {\sl Phys.\ Rev.\ Lett.\/}, 71 (1993) 2537;\\
   K.\ Byrum, CDF Collaboration, Proceedings of the 27th International
     Conference on High Energy Physics, Glasgow, (1994), eds. P.\ J.\ Bussey
     and I.G.~Knowles (Inst. of Physics Publ.) p.989.
\bibitem{bbl2}
   G.\ T.\ Bodwin, E.\ Braaten and G.\ P.\ Lepage,
     {\sl Phys. Rev.\/}, D 46 (1992) R1914.
\bibitem{brfl}
   E.\ Braaten and S.\ Fleming, {\sl Phys.\ Rev.\ Lett.\/}, 74 (1995) 3327.
\bibitem{cho1}
   P.\ Cho and A.\ K.\ Leibovich, {\sl Phys.\ Rev.\/}, D 53 (1996) 150.
\bibitem{flem2}
   N.\ Cacciari and M. Kr\"amer, {\sl Phys.\ Rev.\ Lett.\/}, 76 (1996) 4128;\\
   J.\ Amundson, S.\ Fleming and I.\ Maksymyk, UTTG-10-95, hep-ph/9601298.
\bibitem{ours}
   S.\ Gupta and K.\ Sridhar, TIFR-TH-96-04, hep-ph/9601349, to appear
   in {\sl Phys.\ Rev.\/}, D.
\bibitem{br}
   M.\ Beneke and I.\ Rothstein, {\sl Phys.\ Rev.\/}, D 54 (1996) 2005.
\bibitem{tv}
   W-.K.\ Tang and M.\ V\"anttinen, NORDITA-96/18 P, hep-ph/9603266.
\bibitem{flem1}
   S.\ Fleming and I.\ Maksymyk, MADPH-95-922, hep-ph/9512320.
\bibitem{mangano}
   M.\ Mangano and A.\ Petrelli, {\sl Phys.\ Lett.\/}, B 352 (1995) 445.
\bibitem{cho2}
   P.\ Cho and A.\ K.\ Leibovich, {\sl Phys.\ Rev.\/}, D 53 (1996) 6203.
\bibitem{nason}
   P.\ Nason, S.\ Frixione and G.\ Ridolfi,
   Invited talk at the International Conf.\ on Physics in Collisions,
   Cracow, Poland, Jun.\ 1995 (hep-ph/9510253).
\bibitem{pdflib}
   H.\ Plothow-Besch, {\sl Comp.\ Phys.\ Comm.\/}, 75 (1993) 396.
\bibitem{pp300}
   L.\ Antoniazzi {\sl et al.\/}, {\sl Phys.\ Rev.\/}, D 46 (1992) 4828.
\bibitem{ua6}
   C.\ Morel {\sl et al.\/}, {\sl Phys.\ Lett.\/}, B 252 (1990) 505.
\bibitem{pp550}
   V.\ Abramov {\sl et al.\/}, Fermilab Preprint FERMILAB-PUB-91/62-E. 
\bibitem{pp800}
   M. S.\ Kowitt {\sl et al.\/}, {\sl Phys.\ Rev.\ Lett.\/}, 72 (1994) 1318.
\bibitem{pip125}
   C.\ Akerlof {\sl et al.\/}, {\sl Phys.\ Rev.\/}, D 48 (1993) 5067.
\bibitem{pip515}
   A.\ Gribushin {\sl et al.\/}, {\sl Phys.\ Rev.\/}, D 53 (1996) 4723. 
\bibitem{nde}
   P.\ Nason, S.\ Dawson and R.\ K.\ Ellis, {\sl Nucl.\ Phys.\/}, B 303 (1988) 607.
\bibitem{sld}
   H.\ Fritzsch, {\sl Phys.\ Lett.\/}, B 67 (1977) 217.
   F.\ Halzen, {\sl Phys.\ Lett.\/}, B 69 (1977) 105.
\bibitem{gavai}
   R.\ Gavai {\sl et al.\/}, {\sl Int.\ J.\ Mod.\ Phys.\/}, A 10 (1995) 3043;\\
   J.\ Amundson {\sl et al.\/}, preprint MADPH-96-942, hep-ph/9605295.
\bibitem{ours2}
   S.\ Gupta and K.\ Sridhar, work in progress.
\bibitem{chiexp}
   V.\ Koreshev {\sl et al.\/}, Fermilab Preprint, FERMILAB-PUB-96/093-E. 
\bibitem{na14}
   R.\ Barate {\sl et al.\/}, {\sl Z.\ Phys.\/}, C 33 (1987) 505.
\end{thebibliography}
\end{document}